# DIELECTRICITY AND HARD PHONONS

M. Weger
Racah Institute of Physics, Hebrew University, Jerusalem, Israel

J.I. Birman
Department of Physics, New York City College, Convent Ave. and 138th st. NY

Abstract.

   The maximum value of the superconducting transition temperature $T_c$ due to a phonon-mediated interaction was estimated by Cohen and Anderson in 1972 from *ab-initio* considerations, and found to be about 10 K. McMillan's semi-empirical estimate from 1968 gives a value of about 40 K. We consider these estimates on the basis of subsequent theoretical and experimental evidence, and pay attention in particular to the inhomogeneity of the electron gas. This inhomogeneity gives rise to local-field effects, which are mentioned by Cohen and Anderson, however without an explicit estimate of their effect on $T_c$. They claim that strong local-field effects cause a transition to covalent bonds, which inhibits superconductivity. We consider here strong local-field effects by making use of the inverse dielectric matrix (in reciprocal space), which we review in some detail. We distinguish between the electronic and ionic dielectric matrices, and show that the later can give rise to much stronger local fields, because of the inherent local nature of the ions (in contrast with the delocalised nature of the conduction electrons). Moreover, the ionic dielectric matrix is large only at very low frequencies $\omega$ (below the dispersion frequency of the dielectric constant), therefore the strong local-field effects do not cause the instability leading to the formation of covalent bonds. As a result, the maximum $T_c$ can be much higher than the value estimated by Cohen and Anderson, and by McMillan, and can reach about 200 K in nearly-ferroelectric materials like the high-$T_c$ cuprates.

1) The Impossible Made Common.

Sven Hartmann's spectacular discovery of photon echoes came as a surprise to the "experts" in the field, who were sure that photon echoes cannot occur. Some people argued that the spin-1/2 formalism that accounts for Hahn's spin echoes, cannot be applied to two energy levels of an atom. Others claimed that radiation damping would kill the echo. Erwin Hahn had previously discovered spin-echoes in the induction-field of a magnetic coil, that can be called "virtual photons". Hartmann discovered echoes in the radiation-field of light, *i.e.* due to **real** photons. This observation required a careful and meticulous coordination of the directions of the exciting pulses, *i.e.* their q-values, and the times involved; and, an optically-thin sample, to avoid dephasing effects; and also a very precise alignment of the external magnetic field. In this way, Sven made the impossible – common.

A particular confluence of various factors can give rise to physical properties that are unexpected and counterintuitive. We claim here that high-temperature superconductivity is also an example of a confluence of diverse factors that give rise to a counter-intuitive behavior. Specifically, a dielectric constant is known to shield electrostatic interactions and to reduce the strength of the potential. We claim that under certain special conditions, an ionic dielectric constant can actually enhance the strength of the potential, and in this way enhance the strength of the interaction of electrons with hard phonons (not related to the relatively soft phonons that are responsible for the large dielectric constant). In this way, the superconducting transition temperature $T_c$ arising from the BCS phonon mechanism is increased by a large amount. This phenomenon may take place in the cuprates, in organic superconductors, and in Na-doped $WO_3$, all of which pssess anomalously large dielectric constants. Thus, the impossible becomes almost common.

2) The Maximum Value of $T_c$ in Conventional Superconductors.

A short time after the BCS theory was presented, Cohen and Anderson set out to determine from *ab-initio* considerations the maximum value of $T_c$ possible [1].

They used a "jellium" model, and came to the conclusion that the maximum $T_c$ is about 10 K. Because of the homogeneity of the electron-gas in their model, they find that to maintain stability, the electron-phonon coupling constant $\lambda$ at most equals the bare Coulomb repulsion $\mu$, and superconductivity is possible only because $\mu$ is renormalized to $\mu^* = \mu/[1+ \ln(E_F/\Omega_{phonon})]$ by the Bogolyubov mechanism.

Already at that time there were superconductors with $T_c \approx 20$ K, like NbN, $Nb_3Sn$, *etc*. In the jellium model, local-field corrections are overlooked. McMillan considered such corrections (implicitly), and predicted a maximum $T_c$ of 30-40 K [2]. Because of the local-field corrections, the electron-ion interaction $\lambda$ samples regions in space where the ionic potential is extremely strong (namely the close proximity of the ions). In contrast, the repulsive electron-electron interaction $\mu$ which presses $T_c$ down, involves the average distance between conduction electrons, which is not so small and therefore $\mu$ is smaller than $\lambda$. (Typical values in McMillan's calculation are $\lambda\approx 2$, $\mu\approx 0.5$, $\mu*\approx 0.13$). Cohen and Anderson describe the local-field effects by Umklapp processes, but are not explicit about the maximum $T_c$ possible when they are included. They say, however, that when these effects are very strong, covalent bonding will result, instead of high-temperature superconductivity.

The tendency to covalent bonding will turn the material into a semiconductor or insulator in **most** cases; however, Cohen and Anderson did not claim that there can be no exception to this rule. When the organic metal TTF-TCNQ was discovered, it was found empirically that the electron-phonon coupling in it is large, and can in principle give rise to a high $T_c$. This is because the extreme **inhomogeneity** between the covalent bond, which gives a large electron-phonon Frohlich constant g, and the weak Van-der-Waals bonds, which give rise to a narrow band with a relatively large electronic density-of-states [3]. This situation manifests itself spectacularly in the fullerenes, where $T_c$ is high due to an interaction mediated by **hard** phonons [4]. This inhomogeneity is thus a manifestation of the electronic Umklapp (or local-field effect) discussed by Cohen and Anderson. The initial motivation of Muller to work on the perovskites was also due to the abnormally large electron-phonon coupling of the Cu-O covalent bond [5].

In his estimate for the maximum $T_c$, McMillan used the relationship: $V(q=0)=Z/N(0)$, where $V(q)$ is the electron-ion potential, Z the charge of the ion, and $N(0)$ the bare density of states at the Fermi level. This relationship assumes a homogeneous system [6]. Thus, the inhomogeneity leading to the local-field corrections is not fully taken into account. The limitation of maximum $T_c$ of McMillan thus has a different cause from that due to Cohen and Anderson; it follows from a limitation of the Hopfield parameter $<I^2>N(0)$ in the real systems investigated by McMillan, (where the inhomogeneity is not as large as in the fullerenes, or the cuprates), and not from the stability consideration requiring $\lambda\leq\mu$.

Ginzburg and his associates criticized the work of Cohen and Anderson from *ab-initio* considerations. They showed that the relationship $\lambda\leq\mu$ follows from the relationship for the electronic dielectric constant: $\varepsilon^{-1}(q,\omega\approx 0) \leq 1$, and this relationship is very general, since $\varepsilon(q\approx 0) \geq 1$. However, the relationship for $\varepsilon^{-1}$ can be satisfied also when $\varepsilon$ is **negative**. And, for large values of q ($q\approx k_F$), there is no **general** rule prohibiting a negative $\varepsilon$. Under such conditions, $\lambda$ can be much larger than $\mu$, and $T_c$ can be very much higher [7]. Thus, local-field effects that are so enormous that even the sign of $\varepsilon$ is reversed, are in principle possible for high -q values, *i.e.* locally in r-space.

This approach of Ginzburg was pursued by the group at the Lebedev institute for about 25 years very thoroughly and in great detail [8]. The results are unfortunately disappointing.

3) Superconductors with a Large Dielectric Constant.

An anomalously large dielectric constant is observed in organic metals, cuprates, and Na-doped $WO_3$ . However, unlike the situation in the work of Ginzburg's group, here the enormous dielectric constant is **ionic** (rather than electronic). This dielectric constant can be described by the Lyddane-Sachs-Teller theory of 1941 [9] and is given by:

$$\varepsilon_{ion}(\omega) = \varepsilon_\infty(\omega^2-\omega_{long}^2)/(\omega^2-\omega_{trans}^2) \ . \qquad (1)$$

The dispersive behavior is seen in YBaCuO [10], LaSrCuO [11], and BEDT-TTF$_2$X [12]. $\omega_{trans}$ is the frequency of a low-lying Einstein phonon. This phonon is **not** coupled to the electrons in the "normal" way, as described by Bloch in 1928, Bardeen in 1937, Frohlich and McMillan. This is seen from the extremely sharp dispersion curve [11].

The reason for the absence of normal coupling is that the polarization of these phonons is transverse, *i.e.* in the c-direction, while the electrons move in the ab plane. For this reason it is commonly believed that these phonons have nothing to do with the superconductivity. We believe that this common belief is not correct, but the cause for this is subtle and complex.

From a formal point of view, Birman and Solomon pointed out already in 1982 [13] that group theory makes it possible to describe a close connection between superconductivity, and the phenomena of charge density waves, or ferro and antiferromagnetism, or ferroelectricity – and the last phenomenon is on equal footing with the antiferromagnetism that is "conventionally" believed to be the source of high-temperature superconductivity. We elaborate on this point in a current publication [14].

4) Some Background on Dielectric Response Function and Superconductivity.

Even before the advent of the BCS theory, Landau and Lifshitz [15] appear to have been the first to call attention to the possible relation of a very high static dielectric constant $\varepsilon(0)$ of purely electronic origin and superconductivity in a metal.

A high $\varepsilon (0)$ will require that displacement currents be included in the analysis of the electrodynamics of these materials [16]. We shall return to this point later when we discuss including the ionic contribution to $\varepsilon(\omega)$. The path of expressing the electron-phonon interaction effects by means of the dielectric function $\varepsilon(q,\omega)$ including the wavevector q of momentum transfer, was followed by several authors. For example Kirzhnits *et al* [17] formulated the strong-coupling (Eliashberg) theory of superconductivity using $\varepsilon(q,\omega)$ and the inverse $\varepsilon^{-1}(q,\omega)$, especially in the form of the spectral representations to achieve convenient expressions for $T_c$. Cohen and Anderson [1] based their analysis of maximum $T_c$, and instability of the metallic state, on the non-negative property $\varepsilon(q\approx0,0) \geq 1$ , which was also discussed in [17].

A discussion of the use of the dielectric function ε(q+G,q+G';ω) including "local field" effects was given in chapter 12 of the treatise on superconductivity edited by Parks [18], by M.L. Cohen. In that work the dielectric function is employed in connection with the solution of the Eliashberg equations for weakly coupled, multivalley superconducting semiconductors. The theory was applied to n-SrTiO$_3$, but later work did not support the multivalley model for this material [19]. Further discussion of the dielectric function was given by Ginzburg and Kirzhnits [8]. The "local field" and "non-local" effects take account of rapid field variations (in space) between and near individual ions ("local field" effect), and the longer range of Coulomb interactions (the "non-local" effect), while retardation is signaled by the frequency dependence ω. These effects were discussed in influential papers by Adler [20] and Wiser [21], who discussed the "longitudinal" and "transverse" separation and various asymptotic limits (q,q' → 0, *etc.*) as well as giving practical methods to calculate ε(q,q',ω).

Notice should be taken here of the important work on sum-rules, and the range of validity of the Kramers-Kronig relations which was carried-out by Martin [22] and Kirzhnits [23] at this time. A review paper by Dolgov & Maksimov [24] carried further some possible effects due to local field and strong coupling on the superconducting transition temperatures of a metal. A more recent comprehensive review of the calculations and properties of ε(q,q',ω) is given in the monograph edited by L. V. Keldysh *et al* [25] and in particular the chapter by Dolgov & Maksimov is relevant to the electronic contribution to it.

The emergence of superconductivity in doped SrTiO$_3$, predicted and analyzed by M.L. Cohen [18] required inclusion of the ionic lattice contributions to the total dielectric function. This was carried out by using the non-local (q,ω)-dependent form for both contributions: electronic and lattice ionic. The full local-field effects were not included in the dielectric function used with the Eliashberg equations by Cohen [18].

Turning now to the lattice ionic contribution to ε(q,q',ω), note the early detailed paper of Pick *et al* [26] which connected the dielectric function with the equations of lattice dynamics for the phonon dispersion in simple mono and diatomic crystals.

Concluding this brief overview, we take note of the important work of Hanke [27]. This paper gives a comprehensive review of the microscopic theory starting from the solution of Maxwell's equations, and using a self-consistent field approach to the many-body effects. Both basic description and comparison with the use of an extended bases is given in the calculation of the full ε(q,q',ω). The relation of phonon softening (near-instability) and superconducting phase transition is given, thus extending the earlier considerations of Cohen and Anderson [1].

Despite the several papers cited here on the general response theory, and the importance of using the complete dielectric function, as far as we can determine, the present work is one of the limited number which attempt a concrete calculation of the response function and its application to real-world systems, *i.e.* the high-temperature cuprates.

Returning to the matter of incorporating the displacement current into the electrodynamics of superconductors with a soft lattice (phonon) mode, or "nearly ferroelectric" superconductors, we note that a very simple model of such a system was recently examined [28]. The dielectric function was taken as a sum of the ion (LST-like) contribution of equation (1), plus a "London" contribution from the condensed electrons in the supercurrent due to $\varepsilon(\omega) = \varepsilon_{ion} + \varepsilon_{electron}$ . Solving Maxwell's equations gives novel optical effects (reflection/transmission anomalies), as well as frequency dependent re-entrant phase changes (insulator-superconductor), and also new coupled modes of the system which can be considered as phonon polaritons dressed by the supercurrent, or London electronic supercurrent clothed by phonons. Whichever way one views these new excitations, they illustrate the dramatic effects which can occur (even at the lowest level where local field and non-local effects are ignored !) when the ionic and the electronic dielectric response are taken into account in producing the displacement current.

5) The Dielectric Tensor.

Our objective in this paper is to implement the physical picture with two kinds of contributions (ionic & electronic) to the dielectric function in the complex class of materials typified by the high temperature cuprates, sodium tungsten bronzes, organic superconductors, and related materials.

To do this, we should use the general linear response theory for a medium with an applied "external" electric field $\mathcal{E}(r)$ or potential $V_0(r)$ . The particular form of dielectric response theory that we use, gives rise to the "non-local" dielectric function, which includes "local field effects" described above in paragraphs 2 and 3 of section 4. That is, the electron-ion interaction characterized by the McMillan $\lambda$ , samples in part a short range, inhomogeneous region of strong and rapidly varying potential, and also in part a long-range homogeneous, slowly varying potential as in the usual electron-acoustic phonon interactions. We will formulate this in the dielectric function approach. A brief review of the dielectric tensor will be given here, based on the standard reference [25,27].

The most general linear response of the medium to the applied electric field is:

$$D(r,t) = \iint d^3r' \, dt' \, \varepsilon(r,r',t-t') \, \mathcal{E}(r',t')$$

where time-translation homogeneity was assumed. For a perfect crystal and spatial translational symmetry we must have for the dielectric function $\varepsilon$ :

$$\varepsilon(r,r';\tau) = \varepsilon(r+R_L, r'+R_L; \tau) \qquad (2)$$

where $R_L$ is a crystal lattice vector and $\tau = t - t'$ . Taking the Fourier transform of equation (2) gives:

$$D(q+G;\omega) = \sum_{G'} \varepsilon(q+G,q+G';\omega) \, \mathcal{E}(q+G';\omega) \tag{3}$$

where the non-local dielectric function is defined by:

$$\varepsilon(r,r';\tau) = (1/v_g) \sum_{q}\sum_{G}\sum_{G'} \varepsilon(q+G,q+G';\omega) \cdot \exp\{-i[(q+G)\cdot r - \omega t] + i[(q+G')\cdot r' - \omega t']\} \tag{3a}$$

q is a vector in the first Brillouin zone, and G and G' are reciprocal lattice vectors. The non locality is embedded in the dielectric function which depends on the source point r' and field point r or in reciprocal space, on two reciprocal lattice vectors. If the medium were spatially homogeneous, so that : $\varepsilon(r,r';\tau) = \varepsilon(r-r';\tau)$ then:

$$\varepsilon(q+G,q+G';\omega) = \varepsilon(q+G;\omega) \, \delta_{GG'} . \tag{3b}$$

In this case, the dielectric function is diagonal and no "local field" effect is present. Note that it is the usual practice to refer to cases with G = G' as "diagonal" and G ≠ G' as "non-diagonal". This is consistent with treating $\varepsilon(q+G,q+G';\omega)$ as a matrix in the reciprocal space. The more usual Cartesian tensor indices (ij) as in $\varepsilon_{ij}$ with (i,j) = (x,y,z) are either given explicitly, as for example $\varepsilon_{zz}$, or are simply suppressed. In fact in this work, we are mainly concerned with the zz component $\varepsilon_{zz}$ ; it will be understood that in the absence of explicit indices, zz is intended.

Several authors [20,21,22,23] emphasized that the causal response function to be used is: $\varepsilon^{-1}(q+G,q+G';\omega)$ , and we now briefly recapitulate the argument. We distinguish in our problem the "external" (longitudinal) charge density, which we shall take as the electron or ion unscreened charge density. The corresponding "external potential" is the potential without the dielectric, denoted $V_0(q,\omega)$. In the presence of the dielectric host, a total potential $V(q,\omega)$ arises, incorporating the screening. Now, allowing for the local field effects which originate from the inhomogeneity of the medium we write the general linear relation:

$$V(q+G) = \sum_{G'} \varepsilon^{-1}(q+G,q+G') \, V_0(q+G') \tag{4}$$

where we take $\omega=0$ . In the homogeneous system G = G' so that $V(q) = \varepsilon^{-1}(q) \, V_0(q)$, and $\varepsilon^{-1}(q) = \varepsilon^{-1}(q,q)$ . In particular, note that for G = 0 ,

$$V(q) = \sum_{G'} \varepsilon^{-1}(q,q+G') \, V_0(q+G') \tag{4a}$$

For a non-diagonal dielectric function, $V(q)$ can be bigger than $V_0(q)$ , even if $\varepsilon^{-1}(q,q)$ is smaller than 1 and positive. Obviously, this effect will be important only if the non-diagonal elements of $\varepsilon^{-1}$ are very large.

An important work relating to this aspect, is the calculation of Allender, Bray and Bardeen who suggested that a superconducting temperature of about 70 K should be possible by the excitonic mechanism, in a metal-semiconductor interface (such as Pb-PbTe) [29]. The proposed high $T_c$ depends critically on a highly localized covalent bond in the semiconductor. While the local-field effect was calculated in this work, the inverse dielectric tensor was not calculated explicitly. This work was criticized by Inkson and Anderson [30], who calculated the **electronic** inverse dielectric tensor, and found that its non diagonal elements are not large enough to give the value of $T_c$ calculated by ABB (although the non-diagonal elements of the dielectric tensor itself are large !). The reason why the non-diagonal elements of the inverse tensor are small, is that the valence-band electrons are not sufficiently localized, since localization costs too much kinetic energy (of the order of the width of the valence band). We claim here, that when we deal with the **ionic** inverse dielectric tensor, the non-diagonal elements are large, since the ions are localized by their very nature. Therefore $T_c$ can be high.

We now explore in some detail how a large V(q) will arise from this picture.

6)   The Effective Electron-Ion Potential.

In a 2-D system (characterizing cuprates, organic metals of the (BEDT-TTF)$_2$X type, and the surface superconductivity of the Na doped WO$_3$ [31]), the potential due to a point charge Ze is given by [32]:

$$V_0(q+G) = \frac{2\pi Ze^2}{|q+G|} \quad (5)$$

We are interested in the scattering matrix element between Bloch functions:
$\psi_k(r) = u_k(r)\, e^{ikr}$. For simplicity we assume that $u_k(r)$ does not depend on k, thus:

$$u_k(r) = \sum_G c_G\, e^{iGr}. \quad (6)$$

Thus,

$$\langle k| V_0(q+G)| k+q\rangle = \sum_{G'} \frac{2\pi Ze^2}{|q+G-G'|}\, c_G^*\, c_{G'} \quad (7)$$

And with the ionic dielectric:

$$\langle k | V(q+G) | k+q \rangle = \sum_{G'} \varepsilon_{ion}^{-1}(q+G, q+G') \frac{2\pi Z e^2}{|q+G-G'|} c_G^* c_{G'} \quad (8)$$

This expression still does not take into account the screening by the conduction electrons; in a homogeneous system, this screening is given by:

$$V(q) = \frac{V_0(q)}{1+4\pi\chi_{el}(q)} \quad (9)$$

where $\chi_{el}$ denotes the electronic susceptibility, given by the Lindhard function (we take $\omega=0$ here), given in 2-D by [32]:

$$\chi_{el}(q) = \frac{e^2}{q^2\Omega} \sum_k \frac{f_0(E_k) - f_0(E_{k+q})}{E_{k+q} - E_k} \quad (10)$$

$f_0$ is the Fermi function, which we take at T=0. $E_k$ is the energy, which we take as $k^2/2m^*$ in the first Brillouin zone, and periodically extended. $\Omega$ here is the volume.

Since $\varepsilon_{ion}$ is in the c-direction, and $\chi_{el}$ is in the ab plane, one might expect at first sight that they do not mix, and each acts independently. This is indeed the case for a **uniform** system at q=0. Then, a constant electric field in the ab plane does not produce charges in the bulk (but only at the surface) and as a result it does not induce polarization in the c-direction. But, when q≠0, and/or the system is inhomogeneous, this is not the case. Then the electric field in the ab plane induces charges in the bulk, which induce a component of the polarization in the c-direction, and thus the electronic and ionic susceptibilities interact intimately. When the thickness of the metallic layer d is small compared with the average distance between electrons $r_s a_0^*$, the electron-electron interaction is essentially given by: $e^2/\varepsilon_{ion}|r_1-r_2|$, where $\varepsilon_{ion}$ is the component in the c-direction, and the interaction between the two susceptibilities is very strong.

In an inhomogeneous system,

$$\chi_{el}(q+G) = \frac{e^2}{\Omega} \sum_{k,G'} \varepsilon_{ion}^{-1}(q+G, q+G') \frac{c_G^2 c_{G'}^2}{(q+G-G')^2} \frac{f_0(E_k) - f_0(E_{k+q})}{E_k - E_{k+q}}$$

$$(11)$$

Thus,

(12)
$$\langle k | V_{e-ion}(q+G) | k+q \rangle = \frac{\sum \varepsilon_{ion}^{-1}(q+G,q+G') \frac{2\pi Ze^2}{q+G-G'} c_G^* c_{G'}}{1 + 4\pi \frac{e^2}{\Omega} \sum \varepsilon_{ion}^{-1}(q+G,q+G') \frac{|c_G c_{G'}|^2}{(q+G-G')^2} \frac{f_0(E_{k'}) - f_0(E_{k'+q})}{E_{k'} - E_{k'+q}}}$$

By $V_{e-ion}$ we mean that the electron-ion potential is screened **both** by the ionic dielectric constant, and by the conduction electrons. For our choice of the Bloch functions, the matrix element does not depend on k, but only on q. In order to find the total scattering by crystal-momentum q, we must sum over G; thus we denote:

$$\langle V_{e-ion}(q) \rangle = \sum_G \langle k | V_{e-ion}(q+G) | k+q \rangle \quad . \tag{13}$$

Thus $\langle V_{e-ion}(q) \rangle$ is the effective electron-ion potential that determines the McMillan electron-phonon coupling constant $\lambda$; it plays the role of the quantity denoted $v_q$ by McMillan.

7) Different Cutoffs for Electron-Ion Interaction, and for Response of Electron Gas.

Since the expression for $\langle V_{e-ion}(q) \rangle$ is a little complex, let us make some simplifications to make it more transparent.

The numerator contains terms with G=G', as well as G≠G'. The first can be written as:
$$(2\pi Ze^2/|q|) \sum_G \varepsilon_{ion}^{-1}(q+G,q+G) |c_G|^2 \tag{14}$$

Thus, the ionic dielectric constant is convoluted with $|c_G|^2$, *i.e.* $|c_G|^2$ acts like a "filter" for $\varepsilon_{ion}^{-1}$. Assuming a hydrogenic u(r), with Bohr radius $\tilde{a}_0$, in 2-D: (15)

$$|c_G|^2 = \frac{(2/\tilde{a}_0)^3}{[G^2 + (2/\tilde{a}_0)^2]^{3/2}}$$

and the cutoff of the "filter" is given by $2/\tilde{a}_0$.
For the cuprates, with oxygen 2p orbitals, we can approximate $\tilde{a}_0$ by:

$$\tilde{a}_0 = \frac{n^2 \hbar^2}{me^2 Z_{eff}}$$

With n=2, $Z_{eff}$=4 (the Z=8 nucleus screened by 2 1s and 2 2s electrons). Thus, $\tilde{a}_0 \approx 0.5$ A. Thus the cutoff of about 4 A$^{-1}$ is very high. In contrast, the denominator is the Lindhard function, given by Ando *et al* [32]. It has a near-discontinuity, *i.e.* a sharp cutoff, at $2k_F$.

For the cuprates, $2k_F \approx \sqrt{2}\,\pi/a \approx 1.1$ A$^{-1}$, *i.e.* about 4 times less than the cutoff of the numerator. We illustrate the filter functions in Fig. 1.

The quantity responsible for the very different cutoffs is $k_F a_0\tilde{\,}$, or: $r_s = \sqrt{2}/k_F a_0\tilde{\,}$ (in 2D). $r_s$ characterizes the density of the electron gas. In "normal" metals, $r_s \approx 2\text{–}3$ [33] and the difference between the cutoffs is not very large. However, near the Mott Metal-to-Insulator Transition, $r_s \approx 10$, and the difference in cutoffs is very large.

When the cutoff in the numerator is infinite, then the value of $\varepsilon_{ion}^{-1}$ is the **local** value in r-space (a Fourier transform of a constant in k-space is a $\delta$-function in r-space). Thus the value of $\varepsilon_{ion}^{-1}$ in the numerator is approximately the **local** value at the site of the nucleus (planar oxygen in the case of the cuprates).

When the cutoff in the denominator is smaller than G, then the value of $\varepsilon_{ion}^{-1}$ is essentially $\varepsilon_{ion}^{-1}(q,q)$, *i.e.* the **average** over r-space. Thus, we can write roughly:

$$\langle V_{e-ion}(q)\rangle \cong \frac{2\pi Z e^2 \varepsilon_{local}^{-1}(q)}{q + (2/a_0^*)\varepsilon_{average}^{-1}(q)} \quad (16)$$

($a_0^*$ here is the **band** Bohr radius, $a_0^* = \varepsilon_\infty \hbar^2/m^* e^2$, where m* is the band mass, and $\varepsilon_\infty$ is the value of the background $\varepsilon(q,\omega)$ for $\omega\to\infty$, $q\to 0$; $a_0^*$ is also about 0.5 A).
At q=0, $V_{e-ion}$ is enhanced by a factor: $\varepsilon_{local}^{-1}/\varepsilon_{average}^{-1}$. Naively we might write this factor as: $\varepsilon_{average}/\varepsilon_{local}$, although this is not quite correct since in general: $\langle\varepsilon^{-1}\rangle \neq \langle\varepsilon\rangle^{-1}$ (and in the present case, the difference between these two quantities is large).

In Fig.2 we illustrate $\varepsilon$ and $\varepsilon^{-1}$ as function of position, assuming a cosine dependence of $\varepsilon(r)$ [34]. For a cosine dependence in one dimension, $\langle\varepsilon^{-1}\rangle^{-1} = \sqrt{2}\varepsilon_{local}\langle\varepsilon\rangle$, where $\varepsilon_{local}$ is the minimum value of $\varepsilon(x)$. When $\langle\varepsilon\rangle \approx 30$, then $\langle\varepsilon^{-1}\rangle^{-1} \approx 10$, which is still very large. $\varepsilon_{local}^{-1}$ is about 1, thus the enhancement is by an order-of-magnitude. For a cosine dependence in two dimensions (Fig.3), $\varepsilon(x,y) = \varepsilon_{av} - \Delta\varepsilon\,[\cos(2\pi x/a)+\cos(2\pi y/a)]$, $\langle\varepsilon^{-1}\rangle^{-1} \approx 19$ is nearly twice as large.

Note that $\varepsilon_{ion}^{-1}(q+G,q)$ and $\varepsilon_{ion}^{-1}(q,q+G')$ are the Fourier coefficients of $\varepsilon^{-1}(x)$ in Fig.2. We see that the Fourier coefficients with G or G' not zero are nearly as large as the G=G'=0 coefficients; therefore the non-diagonal elements of $\varepsilon_{ion}^{-1}(q+G,q+G')$ are nearly as large as the diagonal ones.

When instead of Bloch functions concentrated around the atoms we have just plane waves, the cutoff of the "filter" is not so well defined. However, a numerical calculation shows that the effect (eq.16) is nevertheless present.

A microscopic explanation for this behavior is given in ref. [35]. The large dielectric constant is due to the large polarizability of the apex oxygen, as described for example by Kamimura [36] and Rohler [37]. (The hybridization of the apex oxygen orbitals can change easily from the anomalous $sp_z$ to the normal $sp^3$ type). It is probably also due to the motion of the barium atoms in the c-direction; the resonance frequency of the dielectric constant, namely 19 meV [10], is indeed the frequency of the barium vibrations. The apex oxygen sits above the planar copper; thus the regions in the ab plane with large polarizabilities are the copper, and also the empty-hole (below the barium) ones [38]. The planar oxygens do not seem to possess a high polarizability (their vibrations are not seen in the dispersion curve of the dielectric constant). We illustrate this in Fig. 3.

8) Considerations of Stability.

We see that the very large, inhomogeneous dielectric constant can give rise to a large enhancement of the electron-phonon coupling constant $\lambda$, well above the maximum value estimated by McMillan. A very large value of $\lambda$ may be expected to create instabilities of various types. Cohen and Anderson [1] state: "Strong pairing is a dynamic attempt towards bond formation", and suggest an instability toward the formation of covalent bonds. Also, a very large $\lambda$ is known to cause an instability leading to the creation of polarons and bipolarons. This effect was studied intensively by Ranninger [39] and by Mott and Alexandrov [40].

In the present case, $\varepsilon_{ion}$ is a function of $\omega$ with a very low cutoff frequency $\omega_{trans}$ (which is 19 meV in YBaCuO). In $WO_3$, this is the vibration frequency of the heavy tungsten. In the organics, it is about 3.5 meV [12]. We calculated here $<V_{e-ion}(q)>$ for $\omega=0$. For $\omega>\omega_{trans}$, the ionic dielectric constant is no longer large, and $\lambda$ is no longer enhanced. Thus, $\lambda$ is in effect a **renormalized** parameter, with a cutoff frequency $\omega_{trans}$. This frequency is well below the frequency $\Omega$ of the phonons that mediate the attractive electron-electron interaction that gives rise to superconductivity. $\Omega$ in the cuprates is probably about 40 meV [41], and in the organics, about 7-8 meV [42]. In $WO_3$, since $2\Delta/T_c$ has the BCS weak-coupling value [31], the phonons responsible for the pairing are probably very hard (longitudinal oxygen vibrations at about 70 meV). Thus, we deal with two types of phonons, which are physically distinct and play **entirely** different roles. The hard phonon branch $\Omega$ is coupled to the conduction electrons, and thus gives rise to electrical resistivity in the normal state, attractive interaction leading to superconductivity, *etc*. The soft phonon branch $\omega_{trans}$ is **not** coupled to the conduction electrons (in the normal way), but gives rise to a very large dielectric constant $\varepsilon_{cc}$ in the c-direction. Thus the soft phonon branch causes a renormalization of the vertex describing the interaction of the hard phonon with the conduction electrons [43].

In Fig. 4 we illustrate the behavior of the vertex function

$$\Gamma(q,\omega) = \sqrt{\frac{\hbar}{M\Omega}} \langle V_{e-ion}(q) \rangle \qquad (17)$$

as function of q and of ω. The enhancement at small q-values by the dielectric is the effect that we discuss in the present work. At frequencies above $\omega_{trans}$ this effect goes away and the effect of the dielectric medium becomes negligible. Because of the very low cutoff in ω the contribution of the large-λ region to the total energy is small. In particular, the coupling "constant" λ at ω=Ω ,*i.e.* on the "physical sheet", is **weak**. Therefore, there is little pulling of the frequency of the mode Ω . Also, the electrical resistivity is determined by the coupling constant λ at ω=Ω, which is small; therefore, the resistivity (due to the phonon mechanism), is low [44]. The fact that the enhancement of Γ occurs at small-q values, indicates **forward** scattering [45], and this also causes the electrical resistivity to be low. In fact, the effective λ describing the resistivity in the normal state, is two orders smaller than the effective λ that is responsible for the high-$T_c$ superconductivity. This is because in contrast with the phonon pulling, and the resistivity which depend on the value of Γ on the physical sheet, since they are real processes, superconducting pairing is a **virtual** process, which depends on the value of Γ(ω) for values of ω below the superconducting gap Δ (or the temperature T). Thus at very low values of ω, well off the physical sheet. There, Γ is enhanced very much, giving rise to an abnormally large value of $T_c$, without instabilities.

9) Possible Pitfalls.

The effect that we consider is unexpected and counter-intuitive. Therefore, we may easily get trapped in many errors and pitfalls. Some of these are,

(i)     Unconventional Role of Soft Phonons.
Since the work of Bloch in 1928, we know the role of phonons in metals in modulating the potential, as described by the Frohlich Hamiltonian. Here we talk of an entirely different scenario; phonons enter both in the usual fashion (the hard phonons), and also via the Lyddane-Sachs-Teller relation (1) (the soft phonons), which generally applies to insulators and **not** to metals. A metal with a LST ionic dielectric constant appears to be an oxymoron.

(ii)    Cutoffs, k and ω dependencies; Renormalization.
Normally in the theory of superconductivity there are just two cutoffs: The Fermi energy and the Debye frequency. Here there are several cutoffs, some in frequency, and some in momentum. This may cause some confusion. Also, the original BCS theory is non chalant about distinguishing the roles of the momentum k and the frequency ω. Here we must be extremely careful at every stage to verify whether we deal with a k or an ω dependence.

Even in the electron-phonon theory of normal metals, the renormalization function $Z(\omega)$ is quite elusive; because the (phonon) energy is $\omega$ and not $E_k$, it doesn't show up in the normal state resistivity, nor in tunneling measurements, due to a cancellation discussed by Grimvall [46]; for an $E_k$ dependent renormalization, this is not the case. The behavior of the effective mass, as determined from the London penetration depth (Uemura plot), may be elusive as well due to a similar cause.

The London penetration depth $\Lambda$ depends on $n/m^*$. It shows an anomalous decrease as $T_c$ increases (Uemura plot). If n is given by the volume inside the FS (as determined by ARPES or by Shubnikov de Haas oscillations), which stays approximately constant, this means that $m^*$ decreases anomalously.

The work of Uemura, here at the Columbia Physics department, was one of the first important investigations of the *ab-initio* cause of high-temperature superconductivity. The interpretation given was that n increases by doping, while $m^*$ remains constant, and therefore the Bose-Einstein condensation temperature, which is proportional to $n/m^*$, increases; and superconductivity was identified with the BEC. The more recent work that shows that n (the volume inside the FS) stays essentially constant, contradicts this interpretation. The interpretation presented here overcomes this contradiction, since it indicates that $m^*$ decreases. (Technically, this decrease of $m^*$ is attributed to the flat portions of the FS, rather than to the corners where the Van-Hove singularity is [45]. The flat portions are the ones that make the main contribution to the London penetration depth).

When the **electronic** screening is reduced, we may have a scenario similar to Hartree-Fock, where the effective mass $m^*$ **decreases** at the FS [47]. We claim that the electronic screening is reduced by the ionic dielectric constant (eq.11). This effect is related to the renormalization, and is due to the breakdown of the Born Oppenheimer Approximation [48]. Such an effect is admittedly very elusive. The as-yet unaccounted-for $T^2$ dependence of the electrical resistivity in organic metals may also be attributed to this cause [49].

(iii)    Umklapp.

Umklapp processes usually complicate things; here we have not only an effect that depends inherently on strong Umklapp processes, but we must in addition distinguish between electronic Umklapp processes (eq.15) and ionic ones (section 5).

(iv)    Breakdown of the Born-Oppenheimer Approximation.

We deal with an effect that involves a breakdown of the BOA, since we consider a motion of the ions (described by the ionic dielectric function) during the time of interactions of the electrons. Electronic band structure calculations assume the validity of the BOA. Therefore the effect that we consider is not seen at all even in the most careful band structure calculations. Since the results of band structure calculations agree excellently with experiment, we might suspect that we introduce a spurious effect. However, since the effect that we deal with takes place only at very low energies (below $\omega_{trans}$), it does not affect the shape of the Fermi surface, nor the frequency of phonons above this cutoff [50]. Since the phonon $\omega_{trans}$ is not coupled with the electrons (in the normal way), there is no change in the E(k) dispersion even below this cutoff. Therefore a direct observation of the effect proposed here, is elusive. (See subsection (ii)).

(v) Vertex Corrections.

When we deal with the breakdown of the BOA, we usually consider the breakdown of Migdal's theorem, due to vertex corrections [51]. The vertex correction that we consider here is a bubble diagram, and not the more-conventional vertex corrections [43]. The bubble-diagram corrections are usually assumed to be included in the calculation from the start. The fact that they are not, is entirely unexpected.

(vi) Electron Gas Theory.

We deal here with an effect which is outside the conventional Bohm-Pines-Nozieres theory of the electron gas. In that theory, the Debye screening length is the smallest length-scale in the problem (smaller than $k_F^{-1}$, the lattice constant, *etc.*). Here the Debye length is increased by a factor $\sqrt{\varepsilon_{ion}}$ (in 3-D; in 2-D, by $\varepsilon_{ion}$) which is very large. As a result the $\omega=0$ screening length (about 4-8 A) is larger than $k_F^{-1}$, the lattice constant, *etc*. This causes radical modifications in our **conceptual** scenario, and not just modifications of some numerical parameters.

(vii) Dielectric Matrix.

While the dielectric matrix was introduced some 30 years ago, not much work is being done using the full structure of it, and it is somewhat obscure to most people in solid state physics.

(viii) Dimensionality.

There are pitfalls related to the dimensionality. In 3-D, the increase in $V_{e-ion}$ at small q-values is present, but these small q-values occupy only a small volume in phase space [35]. Here we use a 2-D scenario. In 2-D, the potential falls off like $-\ell n\ r$. Since we deal with point charges, we use a potential which falls off like $1/r$. Therefore we got results different from those for plasmas in 2-D and composite layered systems [52]. We deal here with point-charges and a cylindrical (2-D) Fermi surface.

An added complication is, that in 3-D the crossover in the k-dependence of the potential takes place at: $q=q_D/\sqrt{\varepsilon}$ ($q_D$ is the Thomas-Fermi screening parameter), while in 2-D it occurs at: $q=(2/a_0^*)\varepsilon^{-1}$. In the cuprates, these two q-values are rather close, namely around 0.25 $A^{-1}$. Experimentally, stripes are seen with this wavevector [53]. Because of the closeness of the 2-D and 3-D crossovers, we cannot say from the experimental results which scenario is more adequate.

(ix) Effect is Off the Physical Sheet.

We can define the physical sheet as: $\omega=\omega_{phonon}(q)$ where $\omega_{phonon}$ is some phonon branch, acoustic or optic. Here we deal with an effect outside the physical sheet, since $\omega$ is smaller than the frequency of branch $\Omega$ (section 8) at the relevant q-value; *i.e.* we deal with virtual processes.

The distinction between real and virtual processes was of paramount importance already in Sven's discovery of photon echoes, which are due to real photons, in contrast with Hahn's spin echoes, which can be attributed to virtual photons. Thus, the elusive distinction between real and virtual processes pops-up again here.

Because of this distinction, the Hopfield relationship between the normal state resistivity and $T_c$ breaks down by an order-of-magnitude (even after allowing for the forward scattering). This happens both for the magnon-mediated interaction of Pines, and for the phonon mediated interaction. For the former, a renormalized coupling constant was introduced from the start, therefore a Hopfield-type relationship was not expected in the first place. For the later, we are familiar with Eliashberg theory with an unrenormalized coupling constant for 40 years, therefore introducing a renormalized constant (allowing for the proximity of the ferroelectric transition) requires overcoming a severe psychological barrier.

(x)     Eliashberg Equation.

The work with the Eliashberg equation requires the use of a renormalized coupling constant, which was not done before. Since the functions must be analytic in ω (from causality), a non-constant coupling constant involves poles. Simple poles appear in such a calculation as "phonons" (at the frequency of the pole) coupled by some effective coupling constant. The situation here is that the poles are not simple, but second-order poles, whose behavior differs from that of simple poles in an essential way. This requires a careful modification of the computational algorithm. We took the algorithm of Carbotte *et al*. [54], which is already somewhat involved, and introduced the procedure needed to deal with those second-order poles [55].

(xi)    d-wave Symmetry.

The superconducting gap parameter in the cuprates has a d-wave symmetry; this is an argument against the phonon mechanism of pairing. However, for forward scattering with an angle $\Delta\theta/\pi \approx 0.1$, the phonon mechanism (with a weak Coulomb interaction) gives rise to d-wave symmetry [45]. Forward scattering is advocated independently by Kirtley [56].

Here we must also be careful that the condition $\Omega \ll E_F$ necessary for the validity of Migdal's theorem, holds; specifically, $\Omega/E_F$ (which is about 1/25 in the cuprates) must be smaller than $\Delta\theta/\pi$ [57]. We estimate $\Delta\theta/\pi$ to be about 1/10 [45], thus the condition is indeed fulfilled.

Also, the isotope effect in the cuprates is problematic; there is no isotope effect in optimally-doped materials, and there is an isotope effect in under-doped ones. The Coulomb interaction can cause a zero, or even negative isotope effect, as Cohen & Anderson already pointed out. In the present case, the inter-relationship of the phonon-mediated and Coulomb interactions is somewhat more involved; the cutoff of the phonon mediated interaction is (essentially) $\omega_{trans}$, and of the Coulomb interaction it is $\omega_{sf}$, which is treated in detail by Chubukov [58]. When $\omega_{trans} < \omega_{sf}$, there is a "normal" isotope effect, and when $\omega_{trans} > \omega_{sf}$, there is (essentially) a zero isotope effect. $\omega_{sf}$ is estimated to be about 14 meV [58], while $\omega_{trans} \approx 19$ meV. Thus, we are near the crossover region.

(xii)   Double Counting.

In the calculation of "exotic" pairing mechanisms, there is the risk of double-counting of Feynman diagrams. If ε in some of the formulas here were the electronic dielectric constant, we could have double counting. We make sure that ε here is always the

**background** dielectric constant, and categorically exclude the conduction-electron dielectric constant.

The dielectric constant $\varepsilon$ is given by either $\varepsilon = \varepsilon_{ion}^{dressed} + \varepsilon_{el}^{bare} - 1$ or by $\varepsilon = \varepsilon_{ion}^{bare} + \varepsilon_{el}^{dressed} - 1$. If we would take: $\varepsilon_{ion}^{dressed} + \varepsilon_{el}^{dressed} - 1$ (which naively might seem more appropriate), we would double-count. In this work we use $\varepsilon_{ion}^{bare}$ and $\varepsilon_{el}^{dressed}$. This procedure is unconventional, the conventional procedure being to use $\varepsilon_{el}^{bare}$ and $\varepsilon_{ion}^{dressed}$ [ref.33, pp.515-518]; but it is rigorous and there is no double-counting. We explain in detail the reason for adopting this unconventional procedure in ref. [34]; it is due to the limit $\varepsilon_{el} \ll \varepsilon_{ion}$ in our cases.

Also, in the calculation of the vertex renormalization, we have to make sure that the diagrams involved do not appear in the phonon propagator renormalization.

Another aspect of double-counting is rather subtle. We calculate the (microscopic) electron-ion potential V (eqs. 5-13), and make use of the (phenomenological) dielectric constant $\varepsilon_{ion}$ (eqs. 2-13). The dielectric constant induces a polarization, which in turn also gives rise to a potential. Thus, if one considers the potential V, and dielectric constant $\varepsilon_{ion}$ of the **same** ion, there is a double-counting.. In this work we consider a composite material, consisting of metallic regions (such as the $(CuO_2)_n$ planes in the cuprates), and insulating regions in between (apex oxygen, chain copper or bismuth, alkaline earth atoms in the cuprates). The potential V applies **only** to the ions of the metallic region, while the dielectric constant $\varepsilon_{ion}$ applies **only** to the ions of the insulating regions. Therefore there is no double-counting. The calculation in ref. 34 makes this spatial separation very transparent. When we use the measured dielectric constant, we must make sure that the contribution of the metallic regions to it is sufficiently small to be neglected. In the cuprates, this is indeed the case. (We are indebted to Prof. W. Kohn for poiting out this aspect).

(xiii) Coupling between c-axis and ab-plane degrees of freedom.

The c-axis component of the ionic dielectric constant in the cuprates is unusually huge. It is sometimes argued that this degree-of-freedom is decoupled from the dynamics of the conduction electrons, because these move in the ab-plane. This argument is fallacious. Movement of charge along the c-axis, screens out the potential of a charge Q (located in the ab-plane) along the ab-plane; also, movement of charge in the ab-plane, screens-out a potential due to a charge Q along the c-axis. This is evident from elementary electrostatics [34]. We find it hard to understand why such an argument is at all presented.

10) Conclusion.

The Cohen-Anderson stability criterion $\lambda \leq \mu$ and the McMillan rule for the maximum value of the Hopfield parameter $<I^2>N(0)$ give limits to the maximum $T_c$ that are not very much apart (10 K for the Cohen-Anderson criterion, 30 K for McMillan's). They are both overcome by strong Umklapp processes; Cohen and Anderson state this explicitly. Umklapp processes can be electronic, coming from large values of the Fourier coefficients $c_G$ with $G \neq 0$ (eq.12), or ionic, coming from the large non-diagonal elements of $\varepsilon_{ion}^{-1}(q+G, q+G')$. The later were not considered by Cohen & Anderson, and McMillan.

Therefore their presence can lead to a very large increase of a $T_c$ that is due to the phonon mechanism.

Cohen & Anderson state that large (electronic) Umklapp processes lead to the formation of covalent bonds that pre-empt superconductivity. The ionic Umklapp processes have a very low cutoff in ω (the dispersion frequency of $\varepsilon_{ion}$) which is even lower than the frequency of the phonons which are responsible for the pairing; therefore they are not energetic enough to form covalent bonds that prohibit superconductivity. The same argument applies to the Mott-Alexandrov instability leading to the formation of polarons and bipolarons. Thus we are left with a very high $T_c$.

The cause for the breakdown of McMillan's maximum $T_c$ limit is more subtle. McMillan's estimate depends on the sum-rule V(0)=Z/N(0) which applies only in a homogeneous system, and is deeply ingrained in the Bohm-Pines-Nozieres theory of the electron gas. The **large** increase of V(0) above Z/N(0) is due to a reduction of the electronic screening by the ionic screening, which is much larger than the ionic screening *per se,* and thus offsets it by a large amount. This is due to the very large difference in cutoffs (in k-space) for the one-electron ion interaction (cutoff: $2/a_0\tilde{~}$) and the electron-electron interaction (cutoff: $2k_F$). For very large Umklapp terms of the ionic dielectric constant and an electron gas near the Mott transition ($r_s \approx 10$), this gives a large enhancement of the Hopfield parameter $<I^2>N(0)$. Since the effect of ionic dynamics on the electronic properties is outside the Born-Oppenheimer approximation, this effect does not come out of electronic band structure calculations. This effect is also outside the classical Bohm-Pines-Nozieres theory of the electron gas.

While we have to go outside the framework of conventional electronic band structure calculations and electron-gas theory, we **can** use Eliashberg theory, when it is slightly generalized to take into account a phonon propagator normalized by the ionic dielectric function. This was done by Peter *et al* [55], generalizing an algorithm of Carbotte & Marsiglio. This shows the amazing generality of the BCS-Eliashberg theoretical framework.

While the general considerations presented in this conclusion section are not very profound, their detailed implementation opens up a Pandora's box of apparent pitfalls and contradictions to well-established rules. The devil is in the details. We spent years dealing with a large number of details. In this respect we were inspired by Sven's pioneering discovery of photon echoes.

An important recent work of Sven is an echo coming from a single pulse; he addressed it as the sound of a single hand clapping, an expression borrowed from Zen Buddhism. We may make use of this concept in our work as well. Normally, the dielectric function ε is considered to be a function of one q only. In reality, it is a function of two q's. In order to get a high $T_c$ from the phonon mechanism, we need both q's to clap (Umklapp). A single q clapping will only give the maximum $T_c$ predicted by Cohen and Anderson, namely about ten degrees.

Acknowledgements.

This work is a result of close cooperation with M. Peter and J. Gersten. We benefited greatly from discussions with J.P. Carbotte, R.M. Pick, C.S. Ting, H. Ehrenreich, and D. Vanderbilt. The research of M.W. is supported by US-Israel BNF grant 66-00323/3.

Figure Captions.

Fig. 1. The filter functions for the ionic dielectric tensor $\varepsilon_{ion}^{-1}(q+G,q+G')$ for the one-electron scattering by the ion (top) and for the electron-gas response (Lindhard function) (bottom).

Fig. 2. The space-dependence of the dielectric constant $\varepsilon(x)$ and its inverse $\varepsilon^{-1}(x)$ in our model. The averages $\langle\varepsilon\rangle$ and $\langle\varepsilon^{-1}\rangle$ are also indicated.

Fig. 3. Model for the cuprates; the polarizability rests on the apex oxygens, which are situated above the planar coppers, and on the barium atoms, which are above the empty-holes in the $(CuO_2)_n$ lattice. The planar oxygens (denoted O) are not very polarizable. Planes of high polarizability are a distance $a\sqrt{2}$ apart.
  The model Fermi surface is also shown.

Fig. 4. The $\omega$ and q dependencies of the electron – hard phonon vertex function $\Gamma(q,\omega)$. $\Gamma_0$ is the vertex function of the potential $V_0$ (formula 5); $\Gamma$ is the vertex for the potential V of formula (9), and $\tilde{\Gamma}$ the vertex for the potential $V_{e-ion}$ of formula (12).
The vertex function $\tilde{\Gamma}$ is greatly enhanced at small q-values (top) for $\omega$ smaller than $\omega_{trans}$ (bottom).

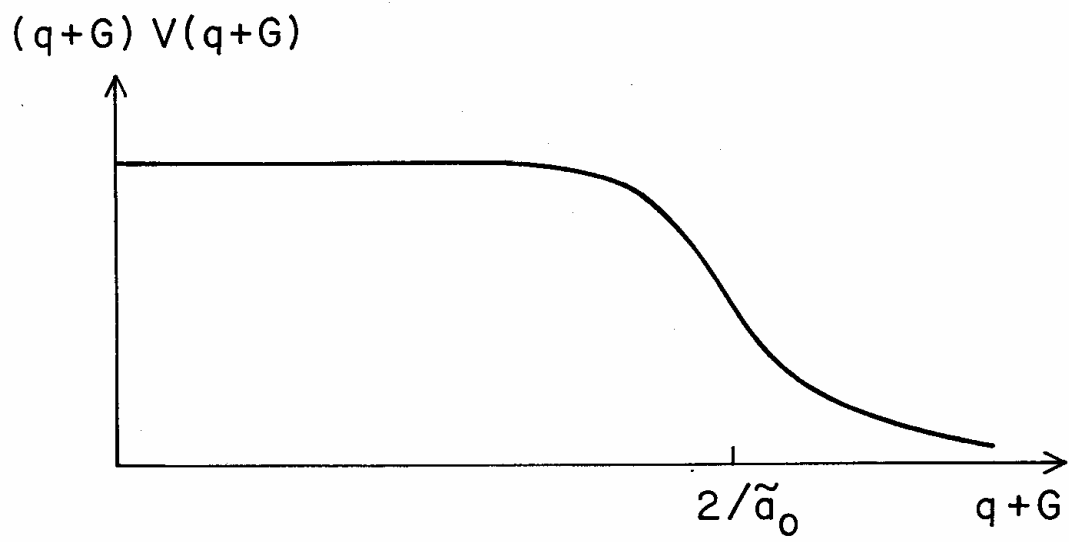
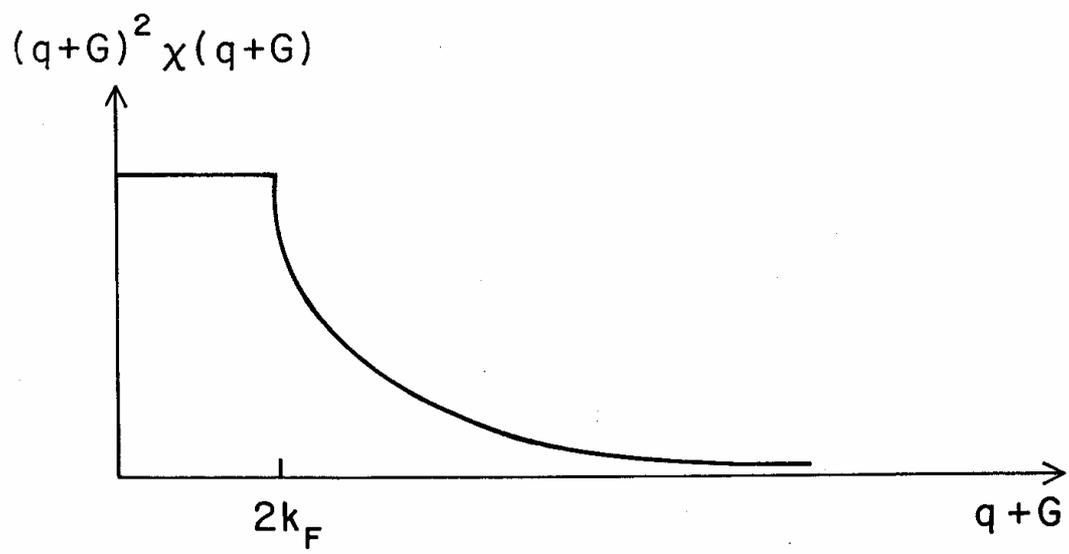

Fig. 1

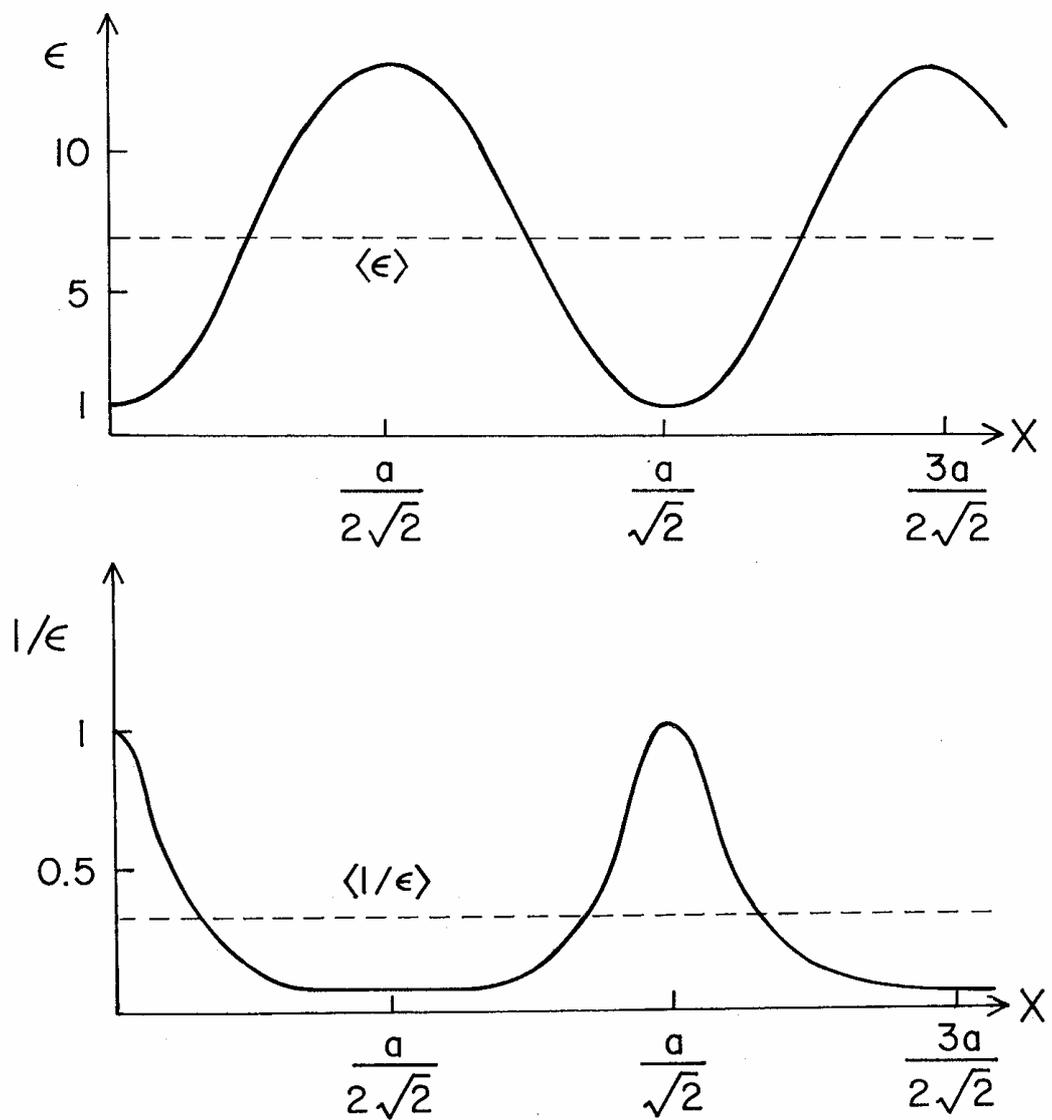

Fig. 2

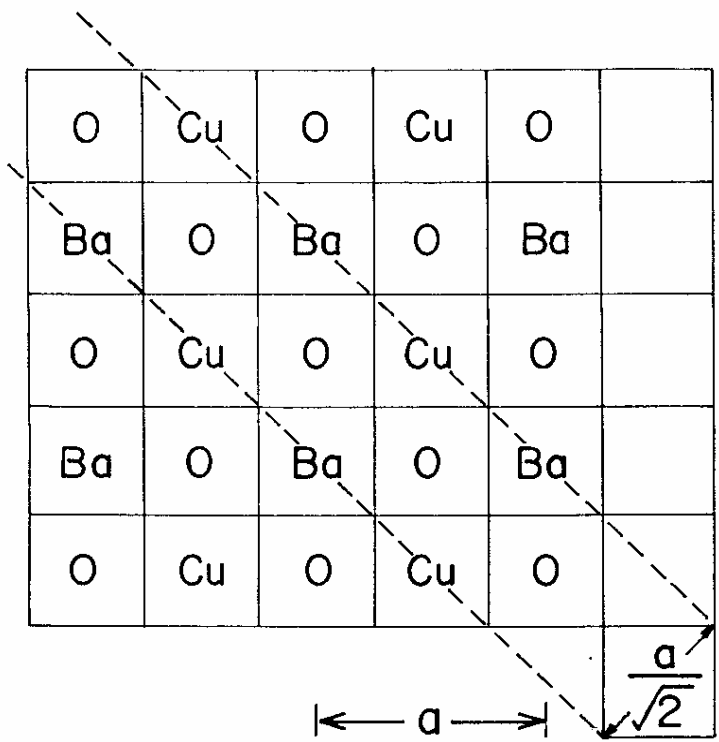

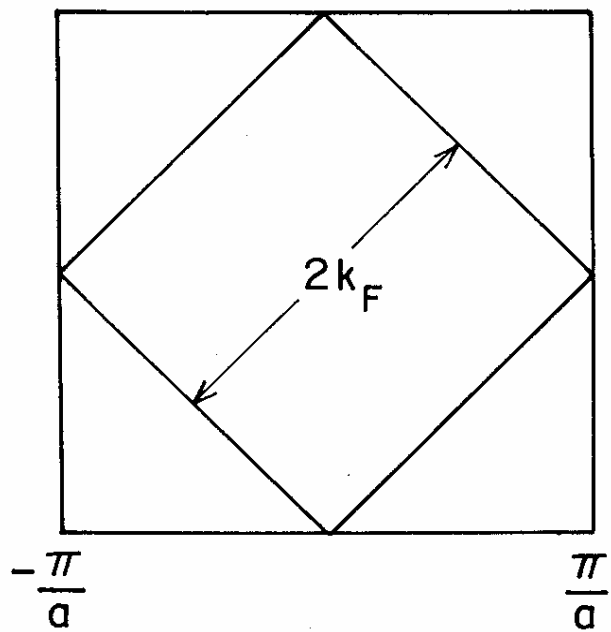

Fig. 3

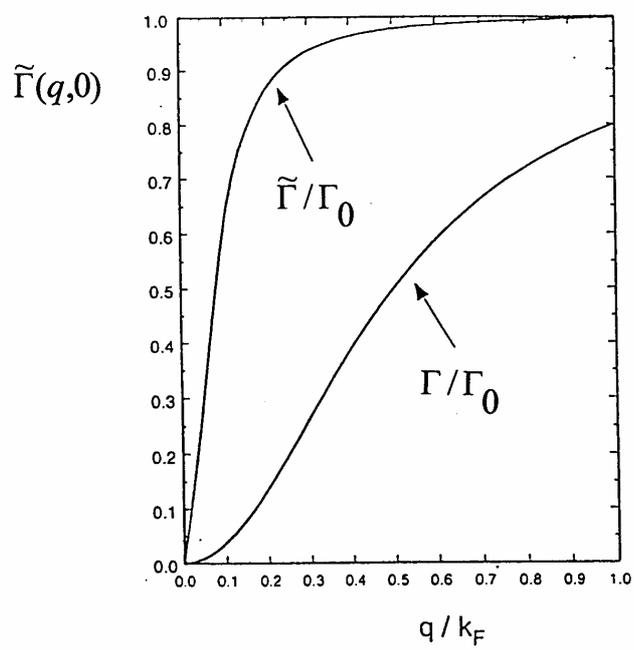
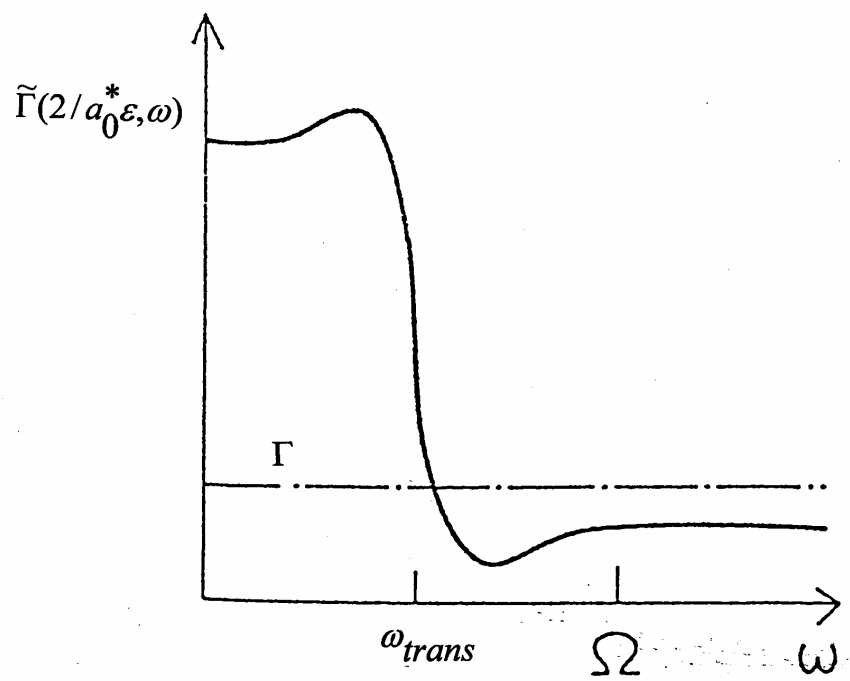

Fig. 4